\documentclass[10pt,conference]{IEEEtran}
\usepackage[style=ieee]{biblatex} 
\usepackage{amsmath}

\usepackage{graphicx}
\usepackage{stfloats}
\usepackage{xcolor}
\usepackage{gensymb}
\usepackage{enumitem}
\usepackage{hyperref}
\begin{document}

\title{Comparing thresholds of $\mathrm{XZZX}$ and rotated surface codes under spatio-temporal correlated noise}
\author{\IEEEauthorblockN{Dhruv Koul}
\IEEEauthorblockA{\textit{Dept. of Physics } \\
\textit{MIT WPU}\\
Pune, India \\
dhruv.koul04@gmail.com}

}
\maketitle 
\begin{abstract}
While independent and identically distributed noise is typically assumed in quantum computing research, complex error models such as correlated and spatio-temporal errors remain underexplored. In this paper, the performance of $\mathrm{XZZX}$ and rotated surface codes is evaluated under these complex noise models. Standard graph-based decoders struggle with the hyperedges generated by correlated errors on bias-tailored codes. To address this, we employ the Minimum Weight Parity Factor (MWPF) decoder, which natively processes syndrome hypergraphs without artificial edge decomposition. Our simulations show that under strong Z-biased circuit-level correlated noise ($\eta=100$), the threshold of the rotated code degrades to $\approx 0.36\%$, while the  $\mathrm{XZZX}$ code drops to $\approx 0.70\%$. Furthermore, a distinct non-monotonic threshold dependence is observed for the  $\mathrm{XZZX}$ code across different bias ratios. While the  $\mathrm{XZZX}$ code effectively absorbs high-density Z-type hyperedges at extreme biases, it exhibits a threshold minimum at moderate biases ($\eta \approx 10$) because transverse X and Y errors actively bridge the diagonal strings, breaking the code's one-dimensional decoupling symmetry. These findings demonstrate that utilizing hypergraph-aware decoders is critical, as the performance of bias-tailored codes is highly sensitive to the geometric structure of spatio-temporal correlated noise.
\end{abstract}

\section{Introduction}

\IEEEPARstart{F}{ault}-tolerant quantum computers are essential for realizing practical quantum computation. Unlike classical computers that primarily face bit-flip errors, quantum computers experience a variety of errors due to decoherence, imperfect gate operations, measurement errors, and environmental noise. To protect quantum information, several error correction techniques are currently being researched. Surface codes are currently one of the leading contenders in this race due to their high error threshold and their ability to operate only using local nearest-neighbor interactions \cite{fowler2012}. To decode errors, decoders must be highly accurate and capable of solving complex decoding graphs in real time without introducing computational bottlenecks.

To reduce the number of physical qubits used to encode logical states in planar surface codes, rotated surface codes were developed \cite{fowler2012}. 
By rotating the lattice by 45$\degree$, the number of physical qubits used is reduced by half, while keeping the code distance and error correction capacity the same.
Although Rotated codes use fewer physical qubits, the $\mathrm{XZZX}$ code \cite{bonilla2021xzzx} was developed to provide better resistance to biased noise.
By changing the stabilizer generators from $\mathrm{XXXX}$ or $\mathrm{ZZZZ}$ to $\mathrm{XZZX}$, this code offers more resistance when dealing with biased noise. 
Recent work by Kam \emph{et al.}~\cite{kam2026spatiotemporal} provides a scalable framework to model physically motivated, spatio-temporal correlated noise, demonstrating that such errors cause severe memory degradation in standard planar surface codes. While standard codes suffer from generalized threshold degradation, this paper demonstrates that spatio-temporal correlations introduce unique, geometry-specific vulnerabilities in  $\mathrm{XZZX}$ and rotated codes by breaking the fundamental decoupling symmetries that bias-tailored codes rely on. Furthermore, by utilizing the Minimum Weight Parity Factor (MWPF) decoder, we isolate these topological vulnerabilities from the artificial algorithmic penalties typically introduced by hyper edge decomposition.

However, the impact of this modeled spatio-temporal noise on resource-optimized variants, such as rotated and $\mathrm{XZZX}$ surface codes, remains unquantified. In practice, on real hardware, noise is not always independent and identically distributed (i.i.d). Errors that affect the qubits can also be correlated in nature \cite{nickerson2019,novais2013,kam2026spatiotemporal}. Understanding the exact limits of topological codes under such complex noise models has become a major focus in recent literature. Recent studies have analyzed the precise effect of multiqubit correlated noise on surface code thresholds \cite{wang2025effect}, while other related work have shown that certain specific correlation symmetries can counterintuitively increase the threshold under certain conditions \cite{wang2025symmetry}. Exact statistical-mechanical mappings have also been developed to map error thresholds strictly under nearest-neighbor correlated noise \cite{wang2025exact}. Despite these analytical advantages, a critical gap remains in the literature. While previous studies have independently evaluated standard codes under spatial crosstalk \cite{wang2025effect}, or bias-tailored codes under purely phenomenological noise \cite{roffe2023bias}, there is no comprehensive evaluation of bias-tailored geometries under a unified spatio-temporal correlated noise model. This work bridges that gap by providing a direct numerical comparison between the $\mathrm{XZZX}$ and rotated surface codes under a dynamic circuit-level detector error model that introduces both temporal decoherence and spatial crosstalk.

Concurrent with our work, Thakur and Van Huele \cite{thakur2026} studied ZX and ZY surface codes under circuit-level biased noise with spatial XX crosstalk. They found that MWPM cannot adequately exploit the doubled syndrome information of tailored codes because correlated errors generate non-matchable hypergraphs, ultimately calling for decoders that can exploit this structure without convergence limitations. Our work directly addresses this gap. By employing the MWPF decoder, we natively handle the hyperedges present in the detector error model. Furthermore, we extend the physical noise model to include continuous temporal correlations alongside spatial crosstalk, evaluating the specific vulnerabilities of the $\mathrm{XZZX}$ and rotated geometries under these conditions.

In the detector error model framework, a detector corresponds to a parity check over a set of measurement outcomes that is deterministic in the absence of noise. That is, under noiseless execution, each detector takes a fixed value across repeated runs, resulting in flips of one or more detectors.
Unlike standard noise models---where a single physical error triggers at most two detectors---correlated errors can trigger three or more. 
In graph-based decoders, these multi-detector events correspond to hyperedges.
Because standard decoders assume noise to be i.i.d, they can process only standard edges, each connecting exactly two detectors and not hyperedges. 
This limitation forces hyperedges to be artificially decomposed. As a result, the decoder loses critical correlation information and miscalculates error probabilities, leading to sub-optimal matchings and a severely degraded threshold. This paper builds on the framework for quantifying this exact threshold degradation specifically for $\mathrm{XZZX}$ and rotated surface codes. By performing circuit-level simulations in STIM \cite{gidney2021stim} and utilizing the MWPF decoder, we provide a direct numerical comparison of these two code families under modeled spatio-temporal correlated noise. Ultimately, while standard graph decoders provide an excellent approximation of maximum likelihood (ML) decoding for standard syndrome graphs, utilizing MWPF ensures that this near-ML decoding accuracy is preserved when processing the complex syndrome hypergraphs generated by modeled hardware noise environments.

The remainder of this paper is organized as follows: Section \ref{sec:background} covers the background on surface codes and their variants. Section \ref{sec:noise_models} covers the noise models and decoding used, followed by Section \ref{sec:simulation} that gives the simulation results. The paper concludes with a discussion and summary in Section \ref{sec:summary}.

\begin{figure*}
    \centering
    \includegraphics[width =\textwidth]{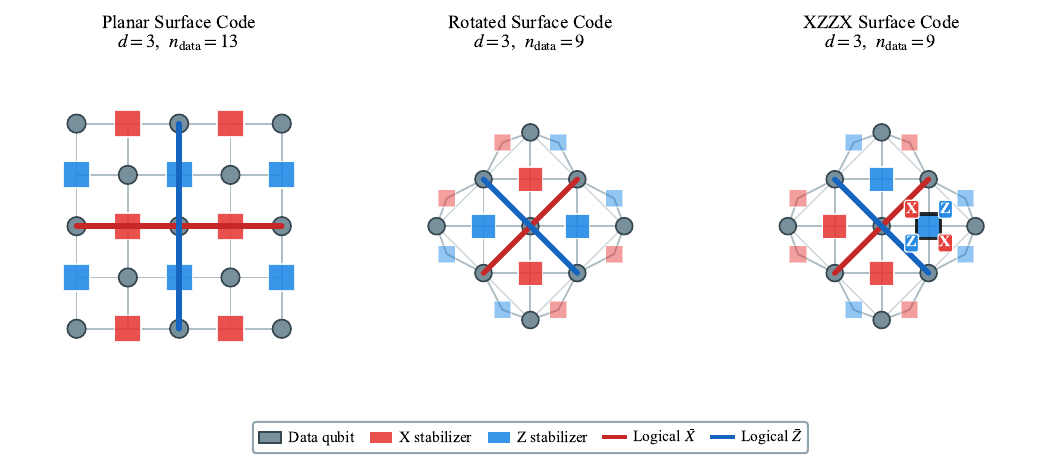}
    \caption{Graphical representation of Planar, Rotated and  $\mathrm{XZZX}$ surface code}
    \label{fig:surface_code}
\end{figure*}

\section{Background}\label{sec:background}
To contextualize the threshold vulnerabilities of bias-tailored codes under correlated noise, we first review the underlying theoretical frameworks of the surface code geometries evaluated in this study.
\subsection{Stabilizer Formalism}
The stabilizer formalism provides a compact framework for describing large quantum codes. Let $\mathcal{P}_n$ denote the \textit{$n$-qubit Pauli group}, defined as
\[
\mathcal{P}_n = \{ \pm 1, \pm i \}\{\mathrm{I}, \mathrm{X}, \mathrm{Y}, \mathrm{Z}\}^{\otimes n},
\]
which consists of all $n$-fold tensor products of single-qubit Pauli operators together with the multiplicative phases $\pm1$ and $\pm i$.
A \textit{stabilizer group} $\mathcal{S}$ is defined as an Abelian subgroup of $\mathcal{P}_n$ satisfying
$
-\mathrm{I} \notin \mathcal{S}.
$
Logical operators on the encoded qubits are represented by Pauli operators $\bar{\mathrm{X}}$ and $\bar{\mathrm{Z}}$ that commute with every element of the stabilizer group but are not themselves elements of $\mathcal{S}$. 
Formally,
\[
\bar{\mathrm{X}}, \bar{\mathrm{Z}} \in N(\mathcal{S}) \setminus \mathcal{S},
\]
where $N(\mathcal{S})$ denotes the normalizer of $\mathcal{S}$ within $\mathcal{P}_n$. These operators preserve the code space while acting non-trivially on the encoded logical qubits.
Error detection in stabilizer codes is performed through \textit{syndrome measurements}. For an error operator $E \in \mathcal{P}_n$ and stabilizer generator $s_i$, the measurement outcome $\lambda_i$ is determined by the commutation relation
\[
\lambda_i =
\begin{cases}
+1, & [E, s_i] = 0 \\
-1, & \{E, s_i\} = 0
\end{cases}
\]
that is, $\lambda_i = +1$ when $E$ commutes with $s_i$, and $\lambda_i = -1$ when $E$ anti-commutes with $s_i$.
Complete syndrome profiles, therefore, map the stabilizers to every detectable error \cite{fowler2012}.
\subsection{Surface Codes} 
While the stabilizer formalism provides mathematical rules for error 
detection, translating them into physical architectures require scalable 
geometric layout. A primary constraint in quantum hardware is the difficulty of engineering long-range interactions between distant qubits. This also places a constraint on the design of any practical quantum error correcting code, that is logical information must be protected using only interactions between physically neighboring qubits.

To achieve large code distances and scalable quantum memory while restricting operations to local nearest-neighbor interactions, topological codes were introduced.  The idea was first proposed by Kitaev \cite{kitaev1997}, where he showed 2 logical qubits can be encoded in a two-dimensional lattice shaped as a toroid with periodic boundary conditions, where the horizontal row encodes the logical $\bar{\mathrm{X}}$ and the vertical lines encode the logical $\bar{\mathrm{Z}}$.
Because implementing periodic boundary conditions is physically impractical on two-dimensional quantum chip architectures, planar surface codes with open boundaries were subsequently developed \cite{fowler2012}. Below are some types of surface codes that are currently being widely used. 
\subsubsection{Rotated Surface Codes}
The lattice is rotated by $45^\circ$ compared to the standard planar construction, allowing a more efficient use of physical qubits. For a code of distance $d$, the rotated surface code \cite{fowler2012} requires $n = d^2$
data qubits and $d^2-1$ ancilla qubits, compared to the $(2d-1)^2$ data qubits required by the unrotated planar code. The stabilizers are divided into two types: $\mathrm{X}$-type and $\mathrm{Z}$-type stabilizers.
Each type has $\frac{d^2 - 1}{2}$ stabilizers. In this structure data qubits lie on the edges of the lattice while ancilla qubits are placed at the center of each face for stabilizer measurements. 
\subsubsection{$\mathrm{XZZX}$ Surface Code}
The $\mathrm{XZZX}$ surface code \cite{bonilla2021xzzx} is a modified variant of the standard surface code which was designed for improved performance under biased noise. Unlike conventional surface codes that use purely $\mathrm{XXXX}$ or $\mathrm{ZZZZ}$ operators on each plaquettes, the $\mathrm{XZZX}$ code applies mixed stabilizer to every face. Specifically, the Pauli operators are arranged diagonally around the vertices to form the $\mathrm{XZZX}$ pattern.
This mixed stabilizer structure allows the code to treat errors in a more symmetric way, particularly when noise is biased toward dephasing.
The $\mathrm{XZZX}$ code effectively maps the two-dimensional error pattern into a set of decoupled one-dimensional repetition codes along the lattice diagonals. 
This significantly improves the error threshold under biased noise conditions \cite{roffe2023bias}. 

However, this one-dimensional symmetry is based on the assumption that errors along different diagonals remain independent.
When spatio-temporal-correlated noise is introduced, errors spread across the lattice, effectively coupling these one-dimensional chains. 
This coupling breaks the decoupling assumption, rendering the $\mathrm{XZZX}$ code particularly vulnerable to cross-chain-correlated errors.

Evaluating exactly how and when this decoupling assumption fails under physical hardware conditions requires a rigorous, circuit-level correlated noise model.
\section{Noise Models and Decoding}\label{sec:noise_models}
Standard evaluations of surface codes frequently assume noise to be independent and identically distributed (i.i.d.). However, to accurately understand their performance on physical hardware, it is critical to evaluate their behavior under correlated, spatio-temporal errors. The primary motivation of this work lies in quantifying this exact threshold degradation large-scale simulations across multiple noise biases. 

The following subsections detail the noise models and the decoding algorithms utilized in this work to natively process and quantify this degradation.
\subsection{Noise}
\subsubsection{Phenomenological noise vs circuit-level noise}
In quantum error correction simulations, noise is typically evaluated at two distinct levels of abstraction: phenomenological and circuit-level noise. Phenomenological noise models assume perfect syndrome extraction and errors occurring only on data qubits and measurement outcomes. On the other hand circuit-level noise is generally more complex to simulate, they inject errors directly into the syndrome extraction, state preparation, entangling gates, idle periods, as well as measurements. It is simulated in order to capture the physical reality of how noise can propagate through circuits. 
\subsubsection{Biased Noise}
Quantum systems often suffer from asymmetric error probabilities, called biased noise. An example of such is a system where pure phase-flip ($\mathrm{Z}$) errors occur at a much higher rate than bit-flip ($\mathrm{X}$) or combined ($\mathrm{Y}$) errors.
To quantify bias, we must first define biased Pauli noise channel. Let $p$ be the total physical error probability and $\eta$ be the bias ratio. The total error rate is the sum of the individual Pauli error probabilities:
$$
p = p_x + p_y + p_z.
$$

In our example the bias ratio $\eta$ defines how much more likely a $\mathrm{Z}$ error is compared to an $\mathrm{X}$ error. Assuming the rates of $\mathrm{X}$ and $\mathrm{Y}$ errors are symmetric ($p_x = p_y$), the bias ratio is defined as:
$$
\eta = \frac{p_z}{p_x}.
$$

Using this definition, we can express the individual Pauli error probabilities in terms of total error rate $p$ and the bias $\eta$. The probabilities for the non-dominant $\mathrm{X}$ and $\mathrm{Y}$ errors are given by:
$$
p_x = p_y = \frac{p}{\eta + 2}.
$$

Similarly, the probability for the dominant error $\mathrm{Z}$ is given by:
$$
p_z = \frac{\eta p}{\eta + 2}.
$$

Under non-biased depolarizing noise, we take $\eta=1$, which gives us an equal probability of $p/3$ for all Pauli errors. To comprehensively map the resilience of these codes, our highly parallelized simulations evaluate threshold shifts across a wide spectrum of biases, specifically $\eta \in \{1, 10, 20, 50, 100, \infty\}$. A highly Z-biased regime, such as $\eta=100$, serves as a primary baseline, allowing us to accurately compare the performance degradation of surface codes under highly biased spatio-temporal conditions.

\subsection{Spatio-Temporal Errors}

While the biased noise model provides a baseline, realistic systems also undergo spatio-temporal-correlated errors \cite{hwang2001correlated, clemens2004quantum}. To accurately quantify threshold degradation, it is critical to evaluate the codes under an additive spatio-temporal penalty framework. If a nominal physical error rate $p$ is simply partitioned into fractional weights across all error mechanisms, the baseline failure rates for standard operations are suppressed. Instead, we define $p$ as the true baseline physical error rate for all standard circuit operations, including entangling gates, state preparation, and measurements.

Spatio-temporal correlations are then introduced as strictly additive overheads on top of this baseline. For temporal noise, continuous memory decoherence is subjected to an overhead factor of $f_{\text{idle}} = 1.0$. For spatial noise, a 50\% crosstalk penalty is distributed equally between adjacent ($f_{\text{adj}} = 0.25$) and diagonal ($f_{\text{diag}} = 0.25$) correlations. This additive model ensures that the performance of the MWPF decoder is measured against a rigorous correlated noise environment layered over an uncompromised standard baseline.

To provide a concrete framework for error injection, the syndrome extraction round is divided into discrete ticks. Based on the standard circuit architecture generated for these simulations, a single extraction round consists of $N_{\text{ticks}} = 7$ discrete clock cycles. Ancilla initialization occurs in the first tick, where State Preparation and Measurement (SPAM) errors are injected at the baseline rate $p$. Entangling gates (CNOT or CZ) are applied in subsequent ticks, immediately followed by two-qubit gate errors at the baseline rate $p$. Finally, ancilla measurements occur, accompanied by measurement errors at rate $p$. During any tick where a qubit is not actively participating in a gate operation, idle errors are injected to simulate continuous decoherence. These errors are injected directly into the syndrome extraction schedule across three distinct physical mechanisms. They are explained in detail below:

\textit{1) Continuous Memory Decoherence:} In physical quantum systems exhibiting heavily biased noise, qubits undergo continuous phase damping ($T_2$) when idle[cite: 25]. Over a continuous time interval $t$, the probability of a physical dephasing error grows exponentially as \cite{nielsen2010quantum}:
\begin{equation}
p(t) = \frac{1}{2} \left( 1 - e^{-t/T_2} \right),
\end{equation}
where $T_2$ represents the characteristic dephasing time[cite: 25, 26, 27]. To model this continuous time evolution within a discrete circuit simulator, independent single-qubit errors are injected at every time-step of the syndrome extraction schedule. To ensure the idle error rate per round mathematically compounds to the target overhead without exceeding bounds, the probability per discrete time-step is calculated using geometric scaling \cite{gidney2021stim, acharya2022suppressing}:
\begin{equation}
p_{\text{tick}} = 1 - (1 - f_{\text{idle}} \cdot p)^{1/N_{\text{ticks}}},
\end{equation}
where $N_{\text{ticks}} = 7$ is the total number of discrete clock cycles per syndrome extraction round and $f_{\text{idle}} = 1.0$ is the temporal penalty factor.

\textit{2) Faulty Entanglement Gates and Spatial Crosstalk:} Two-qubit entangling gates are a major source of error in surface code architectures. In our additive framework, the standard baseline error following each entanglement operation is injected at probability $p$. To capture spatial correlations, we inject additional correlated errors on adjacent and diagonal qubits with probabilities $p_{\text{adj}} = f_{\text{adj}} \cdot p$ and $p_{\text{diag}} = f_{\text{diag}} \cdot p$, respectively. This allows for an accurate representation of the physical hook errors and cross-talk that originate from the gate mechanism and dense physical layouts.

\textit{3) Organic Hyperedge Formation:} Rather than artificially constructing joint probability distributions, correlations emerge deterministically through Pauli error propagation. For instance, consider a continuous memory $\mathrm{Z}$ fault occurring on a data qubit mid-schedule. In subsequent ticks, if this data qubit acts as the control for a CNOT gate targeting an $\mathrm{X}$-measure ancilla, the $\mathrm{Z}$ error commutes with the control and propagates to the target $\mathrm{X}$-ancilla, flipping its measurement outcome. Furthermore, the $\mathrm{Z}$ error remains on the data qubit, where it will flip the subsequent round's $\mathrm{X}$-ancilla measurements. This single propagating fault simultaneously corrupts the spatial data state and flips multiple temporal measurement outcomes. 

This organic circuit-level propagation deterministically generates complex spatio-temporal hyperedges within the detector error model. A detector error model maps each physical error mechanism $e$ to the set of detection events whose outcomes it flips. In the simplest case, an independent error mechanism flips exactly two detection events ($D_e = \{d_i, d_j\}$), corresponding to a standard edge. However, under realistic circuit-level noise, an error may propagate and flip more than two detection events simultaneously; these multi-detection events correspond to hyperedges.

The noise model parameters were chosen to reflect error scaling of superconducting hardwares \cite{google2023suppressing}. The temporal penalty factor $f_{\text{idle}} = 1.0$ assumes that the probability of an idle decoherence fault during a standard gate duration is roughly equivalent to the baseline two-qubit gate error rate itself, a common characteristic in architectures where $T_1/T_2$ times dictate operation speeds. The spatial crosstalk factors of $f_{\text{adj}} = f_{\text{diag}} = 0.25$ represent a 25\% parasitic correlation relative to the baseline error rate. This scales appropriately with observed capacitive crosstalk and microwave pulse leakage to neighboring frequency-multiplexed qubits during simultaneous entanglement operations in dense planar lattices \cite{zhao2022quantum, parrado2021crosstalk, zhou2025surface}. It is important to note that these values ($f_{\text{idle}}$, $f_{\text{adj}}$, $f_{\text{diag}}$) are selected to represent a physically motivated, controlled correlated-noise environment for numerical evaluation, rather than serving as exact empirical measurements of a specific superconducting quantum processor.

\begin{table}[h]
\centering
\caption{Active Error Channels per Simulation Scenario}
\label{tab:noise_models}
\resizebox{\columnwidth}{!}{%
\begin{tabular}{|l|c|c|c|c|}
\hline
\textbf{Error Mechanism} & \textbf{Phenom. Baseline} & \textbf{Circuit Baseline} & \textbf{Phenom. ST} & \textbf{Circuit ST} \\ 
 & (Figure 3) & (Figure 4) & (Figure 5) & (Figure 6) \\ \hline
Data Qubit \& Readout Errors & Active & Active & Active & Active \\ \hline
Idle Decoherence & Inactive & Inactive & Active & Active \\ \hline
Spatial Crosstalk & Inactive & Inactive & Active & Active \\ \hline
State Preparation Errors & Inactive & Active & Inactive & Active \\ \hline
Two-Qubit Gate Errors & Inactive & Active & Inactive & Active \\ \hline
\end{tabular}%
}
\end{table}

\begin{figure*}[h!]
    \centering
    \includegraphics[width=\textwidth]{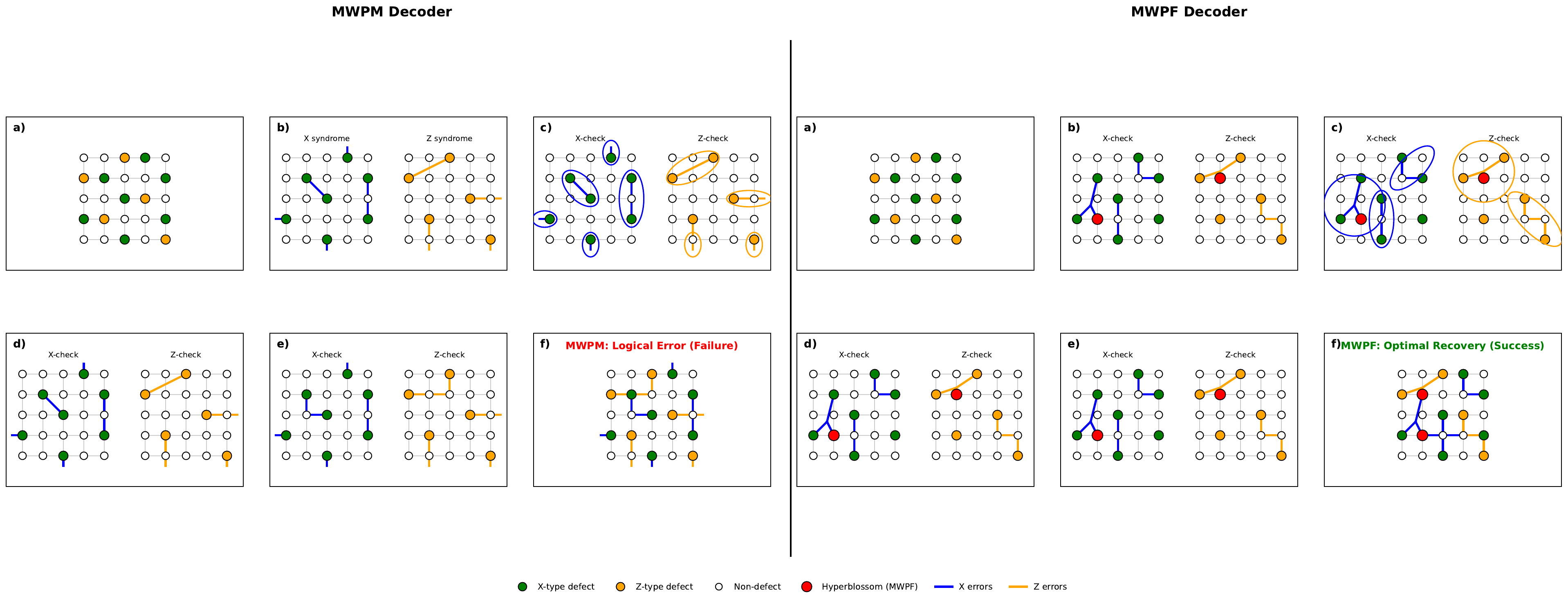}
    \caption{Step-by-step comparison of the MWPM and MWPF decoding procedures applied to an identical syndrome. (a) The initial syndrome, with green and orange nodes representing triggered X-type and Z-type detection events, respectively. (b) The decoding graph constructed from the syndrome: MWPM builds a standard graph of pairwise edges, while MWPF constructs a hypergraph that natively represents multi-detector correlations. Red nodes, called \textit{hyperblossoms}, mark the meeting points where three or more error chains converge, in standard graph decoders these multi-way junctions cannot be represented and must be artificially split into pairs, hence losing the correlation information. MWPF instead treats them as a single unit, preserving the joint structure of the correlated error. (c) Candidate clusterings identified by each decoder, highlighted by ellipses. (d) The selected minimum-weight solution: MWPM commits to a 
    perfect matching over standard edges, while MWPF selects a minimum-weight parity factor over the full hypergraph. (e) The recovered error chains implied by each decoding decision. (f) The final outcome: 
    MWPM produces a logically incorrect recovery due to information lost during hyperedge decomposition, while MWPF produces the correct recovery in this example by preserving the joint probability structure of correlated 
    errors.}
    \label{fig:placeholder}
\end{figure*}

\subsection{Minimum Weight Parity Factor Decoding}

While standard graph-based decoders are highly effective for independent and identically distributed (i.i.d.) noise, they face fundamental limitations when encountering correlated spatio-temporal noise. Because traditional pairwise matching is restricted to a standard graph-based structure, it can only process edges that connect a maximum of two vertices. When a spatio-temporal error mechanism triggers a hyperedge (three or more detection events), these decoders must artificially decompose this hyperedge into standard edges. This decomposition discards the joint probability information of the correlation, leading to sub-optimal matching and a significant degradation in the logical error threshold.

To overcome the limitations of edge decomposition in standard graph based decoders, this work utilizes the MWPF decoder \cite{wu2025mwpf}. Instead of reducing the DEM to a standard graph, the MWPF operates directly on the hypergraph representation of the syndrome. For a given set of triggered detectors, the MWPF algorithm searches for a parity factor---a subgraph of the hypergraph where the degree of each vertex matches the observed syndrome parity.

Mathematically it can be represented as a Maximum Likelihood (MLE) problem, where the goal is to identify the most probable error trajectory that is consistent with the observed syndrome. Let $H = (V, E)$ be the syndrome hypergraph, where $V$ is the set of all detectors and $E$ is the set of hyperedges representing physical error mechanisms. For a subset of triggered detectors $D \subseteq V$, the decoder assigns a binary variable $x_e \in \{0, 1\}$ to each hyperedge $e$, where $x_e = 1$ indicates that the corresponding error has occurred, and $x_e = 0$ indicates it did not. To find the maximum likelihood solution, we minimize the total weight:
$$
\min \sum_{e \in E} w_e x_e,
$$
subject to the vertex parity constraints:
$$
\sum_{e \in E \mid v \in e} x_e \equiv 1 \pmod 2, \quad \forall v \in D,
$$
$$
\sum_{e \in E \mid v \in e} x_e \equiv 0 \pmod 2, \quad \forall v \in V \setminus D,
$$
where the weight of each hyperedge is defined by its log-likelihood, that is:
$$
w_e = \log\left(\frac{1-p_e}{p_e}\right),
$$
where $p_e$ is the physical probability of the error mechanism. By minimizing the sum of these weights, the algorithm effectively maximizes the overall probability of the error trajectory. The resulting subgraph, known as a parity factor, satisfies the specific parity requirements of each vertex in the syndrome hypergraph.

MWPF preserves the joint structure of correlated physical error mechanisms ensuring that the decoding process does not break the topological symmetries of the underlying code, allowing for an unbiased evaluation of the code under correlated noise. Crucially, despite the added structural complexity of processing hyperedges, the MWPF algorithm achieves an average time complexity of $O(|V|)$, where $|V|$ is the number of vertices in the hypergraph.

\section{Simulation and results}
\label{sec:simulation}

\begin{figure*}[h!]
    \centering
    \includegraphics[width=\textwidth]{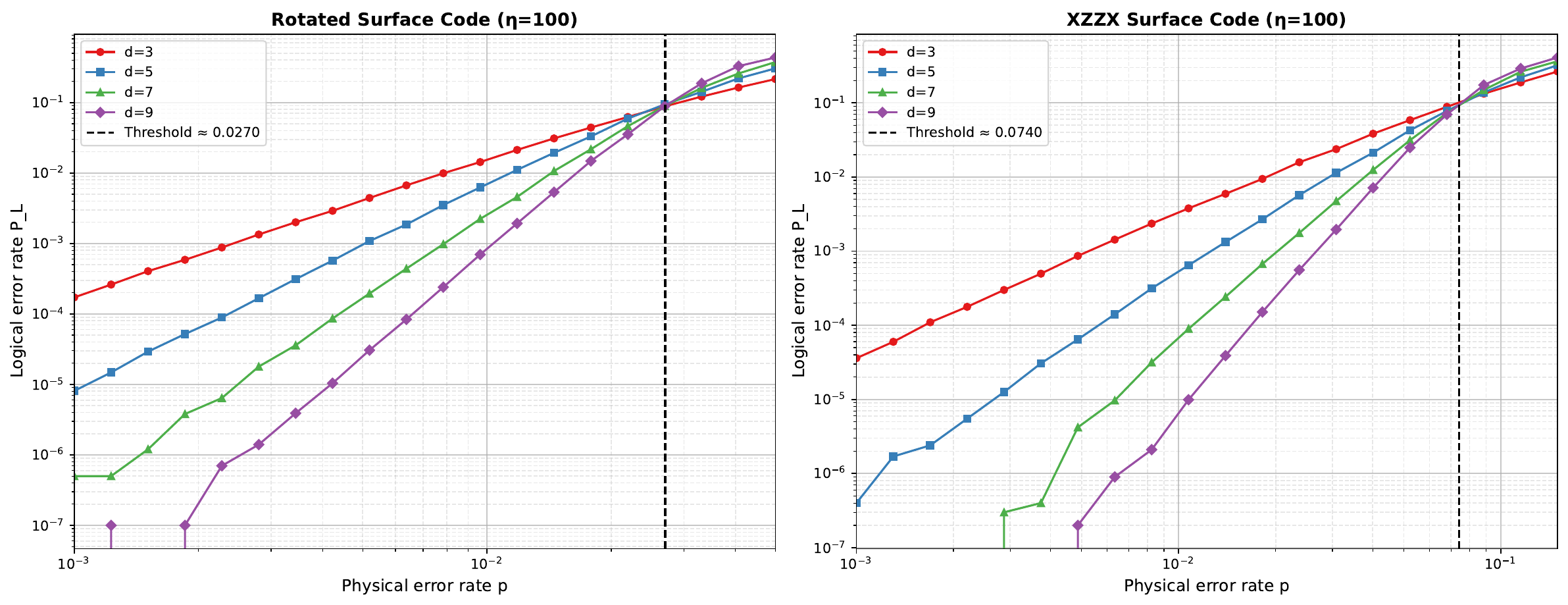}
    \caption{Logical error rate vs.\ physical error rate under strongly Z-biased phenomenological noise ($\eta = 100$) for (left) Rotated Surface Code and (right) $\mathrm{XZZX}$ Surface Code, decoded using MWPF. Under this pure bias baseline, the $\mathrm{XZZX}$ code achieves a threshold of ${\approx}7.40\%$, nearly $2.7\times$ higher than the rotated code's threshold of ${\approx}2.70\%$, showcasing its high resilience to pure dephasing errors.}
    \label{fig:plot1_biased_phenom}
\end{figure*}

\begin{figure*}[h!]
    \centering
    \includegraphics[width=\textwidth]{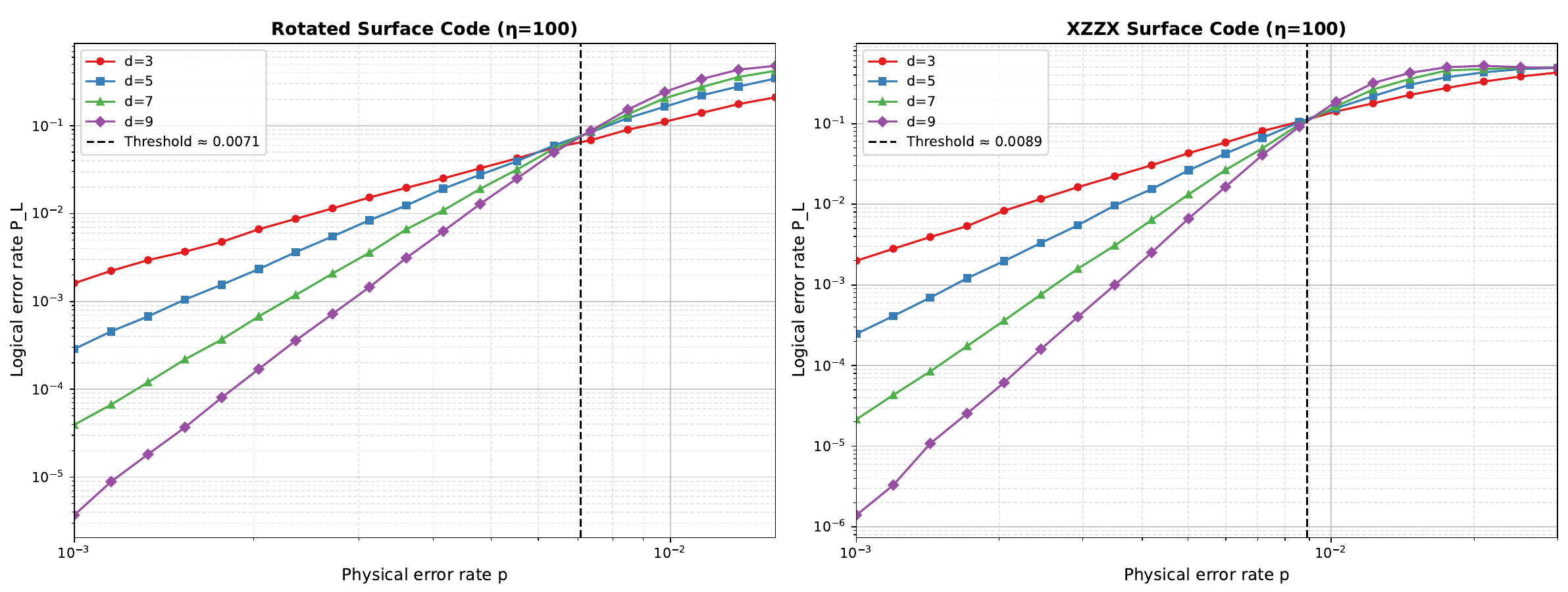}
    \caption{Logical error rate vs.\ physical error rate under strongly Z-biased circuit-level noise ($\eta = 100$) for (left) Rotated Surface Code and (right) $\mathrm{XZZX}$ Surface Code, decoded using MWPF. Introducing circuit-level error propagation yields a threshold of ${\approx}0.89\%$ for the $\mathrm{XZZX}$ code and ${\approx}0.71\%$ for the rotated code, demonstrating that the $\mathrm{XZZX}$ code maintains a topological advantage under pure bias.}
    \label{fig:plot2_biased_circuit}
\end{figure*}

\begin{figure*}[h!]
    \centering
    \includegraphics[width=\textwidth]{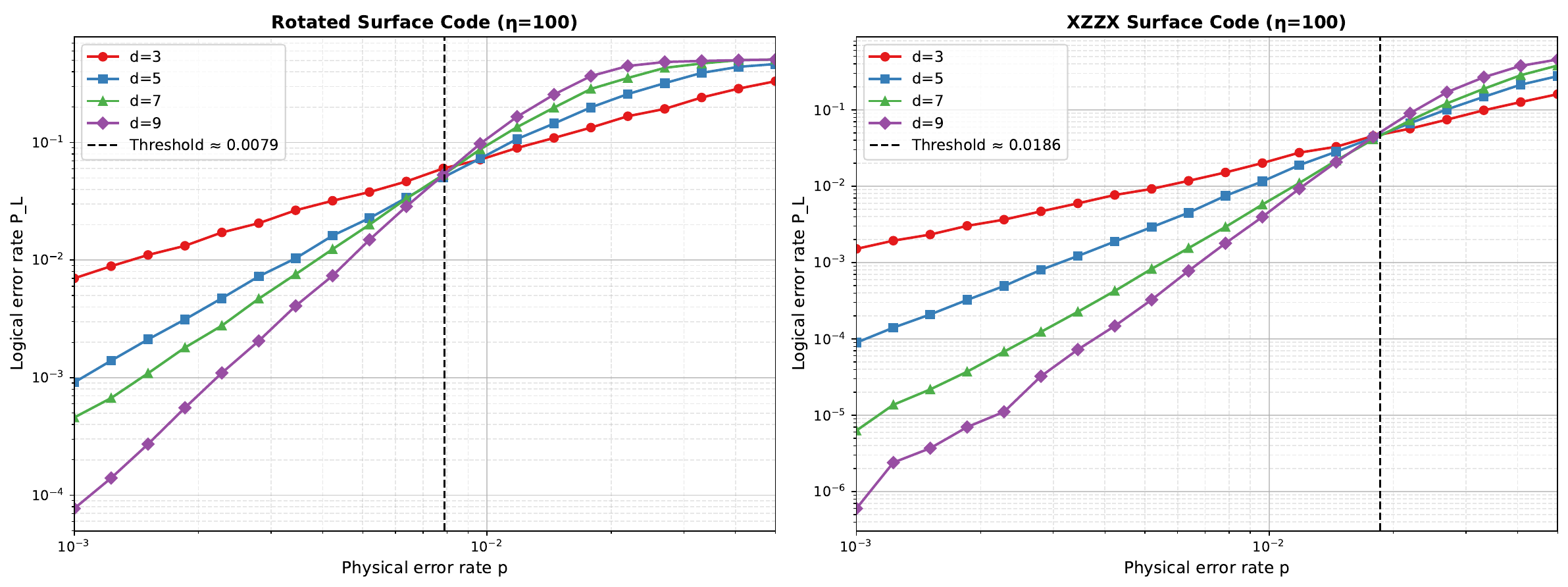}
     \caption{Logical error rate vs.\ physical error rate under Z-biased phenomenological spatio-temporal correlated noise ($\eta = 100$) for (left) Rotated Surface Code and (right) $\mathrm{XZZX}$ Surface Code, decoded using MWPF. The introduction of correlated errors degrades the $\mathrm{XZZX}$ threshold to ${\approx}1.86\%$, while the rotated code drops to ${\approx}0.79\%$, illustrating the vulnerability of decoupled 1D chains to correlated noise.}
    \label{fig:plot3_phenom_st}
\end{figure*}

\begin{figure*}[h!]
    \centering
    \includegraphics[width=\textwidth]{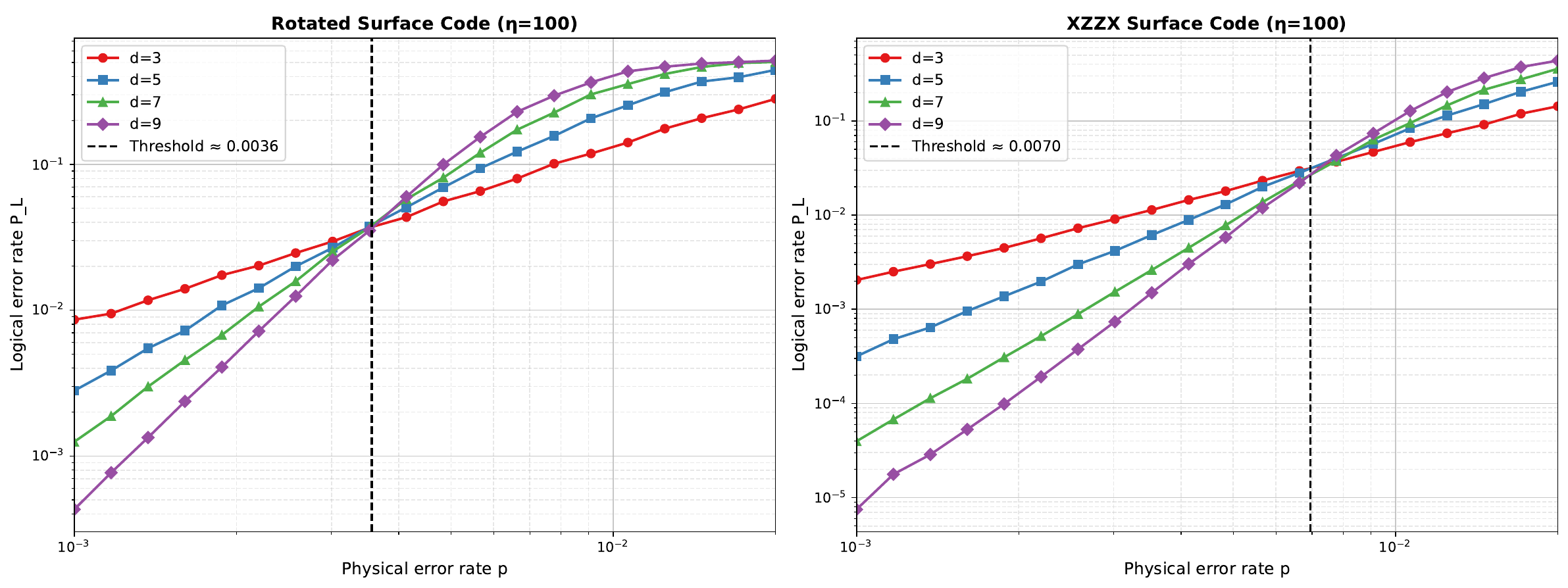}
    \caption{Logical error rate vs.\ physical error rate under Z-biased circuit-level spatio-temporal correlated noise ($\eta = 100$) for (left) Rotated Surface Code and (right) $\mathrm{XZZX}$ Surface Code, decoded using MWPF. The $\mathrm{XZZX}$ code achieves a threshold of ${\approx}0.70\%$, outperforming the rotated code's threshold of ${\approx}0.36\%$. This demonstrates the severe threshold degradation of bias-tailored codes when decoupled 1D chains are disrupted by modeled spatio-temporal correlations.}
    \label{fig:plot4_circuit_st}
\end{figure*}

\begin{figure*}[h!]
    \centering
    \includegraphics[width=\textwidth]{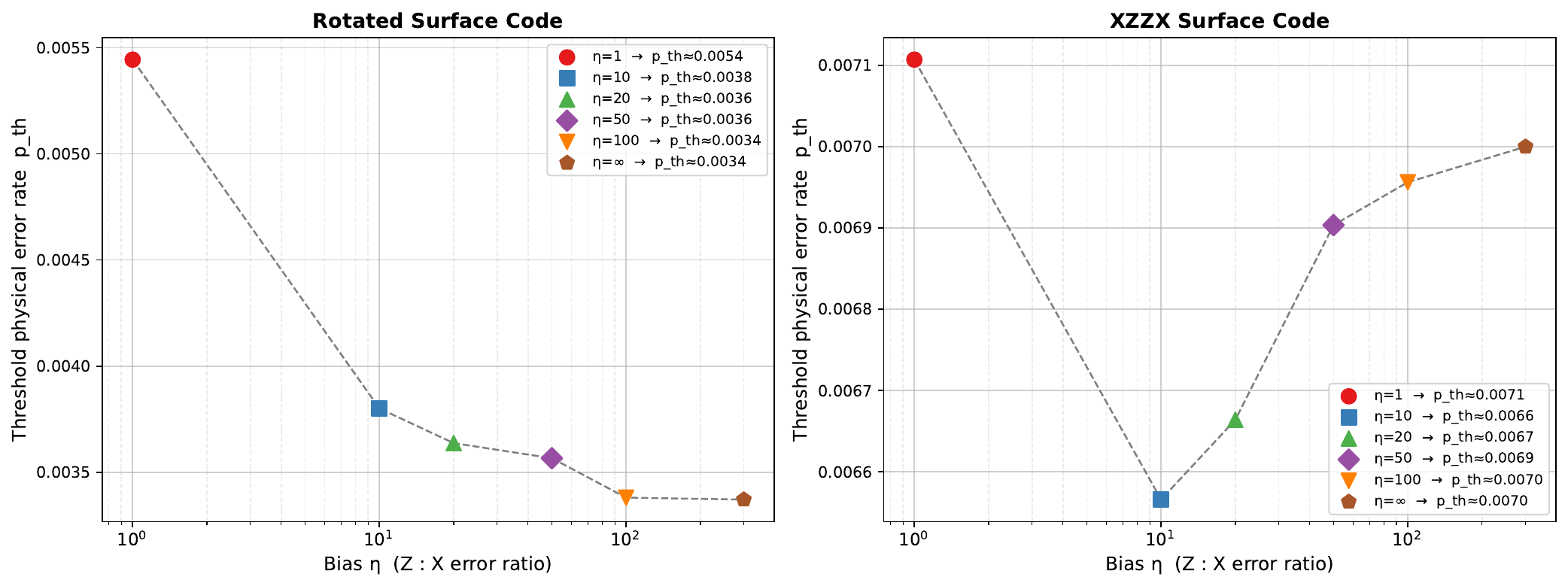}
    \caption{Threshold physical error rate ($p_{th}$) as a function of the Z-bias ratio ($\eta$) for Rotated (left) and  $\mathrm{XZZX}$ (right) surface codes. Simulations were conducted under the modeled circuit-level spatio-temporal correlated noise and decoded using the MWPF decoder across a bias spectrum of $\eta \in \{1, 10, 20, 50, 100, \infty\}$. The rotated surface code exhibits a monotonic degradation in threshold, decreasing from $\approx0.54\%$ under standard depolarizing noise ($\eta=1$) to an asymptotic limit of $\approx0.34\%$ under pure dephasing ($\eta=\infty$). In contrast, the  $\mathrm{XZZX}$ surface code demonstrates a distinct non-monotonic dependence; its threshold drops from $\approx0.71\%$ at $\eta=1$ to a local minimum of $\approx0.66\%$ near $\eta=10$, before recovering asymptotically to $\approx0.70\%$ as $\eta \to \infty$. This illustrates the complex interplay between spatio-temporal hyperedge formation and the noise-tailored stabilizer geometry across differing noise regimes.}
    \label{fig:threshold_vs_eta}
\end{figure*}

\subsection{Simulation Parameters}
To evaluate code performance under hyperedges as described in Section III, large-scale circuit-level simulations were conducted. Processing syndrome hypergraphs with the MWPF decoder isolates true topological resilience, avoiding the artificial accuracy drops due to edge decomposition. To optimize runtime, simulations are bound at either 10,000,000 shots or 2,000 errors. This error bound ensures consistent statistical precision, as relative uncertainty scales as $1/\sqrt{\text{errors}}$. The core parameters are:

\begin{enumerate}[label=(\alph*)]
    \item Code distance ($d$):$\{3, 5, 7, 9\}$
    \item Max Shots: $10,000,000$
    \item Max Errors: $2,000$
    \item Round factor: Same as code distance
    \item Data points per curve: $20$
    \item Physical error rate ($p$): $p \in [0.001, p_{\max}]$, where $p_{\max} \in [0.050, 0.10]$. It is adjusted according to the encoder and noise model to tightly bound their respective threshold crossings.
\end{enumerate}
To determine the logical error threshold $p_{th}$, we identify the scale-invariant crossing point of the logical error rate curves for distances $d \in \{3, 5, 7, 9\}$. The crossing region is isolated, and the critical point is extracted using a polynomial interpolation of the data points nearest to the intersection. It should be noted that all reported $p_{th}$ values are numerically estimated thresholds from finite-distance simulations rather than exact analytical limits. Distances $d \in \{3, 5, 7, 9\}$ were selected to accurately capture the finite-size crossing behavior while keeping the computational cost of processing dense multi-basis hypergraphs manageable.

\subsection{Topological Resilience of Rotated Codes}

Under strongly Z-biased noise conditions ($\eta=100$), when decoded with the MWPF decoder, the rotated surface code demonstrates robust fault-tolerant behavior against spatio-temporal error propagation. A threshold of $p_{th} \approx 0.0036$ was observed under circuit-level spatio-temporal noise, alongside a threshold of $p_{th} \approx 0.0079$ for phenomenological spatio-temporal noise. 

To accurately quantify this performance degradation under bias, these results are compared against the respective baseline thresholds of the rotated code under the same pure Z-bias ($\eta=100$) without spatio-temporal correlations. In those baseline scenarios, the rotated code achieved a threshold of approximately $0.0270$ for phenomenological noise and $0.0071$ for circuit-level noise. The transition to the realistic spatio-temporal model introduces a substantial degradation, reducing the circuit-level threshold by approximately 49\% (dropping from $0.0071$ to $0.0036$). This shift is a direct physical consequence of the organic hyperedge formation previously described. Because continuous dephasing and faulty entangling operations generate correlated detections that span multiple syndrome rounds, the overall complexity of the syndrome hypergraph is significantly increased. 

However, despite this complexity and the extreme noise bias, the rotated code remains highly resilient. Unlike bias-tailored codes, rotated codes natively rely on a symmetric two-dimensional topological structure to protect logical information. Consequently, while the spatio-temporal error hyperedges lower the physical error limit the code can tolerate, they do not fundamentally break the underlying two-dimensional topological symmetry, allowing the rotated code to retain a highly practical and stable threshold even under modeled spatio-temporal correlated noise conditions.

\subsection{Symmetry Breaking in $\mathrm{XZZX}$ Codes}

\begin{figure*}[t]
    \centering
    \includegraphics[width=\textwidth]{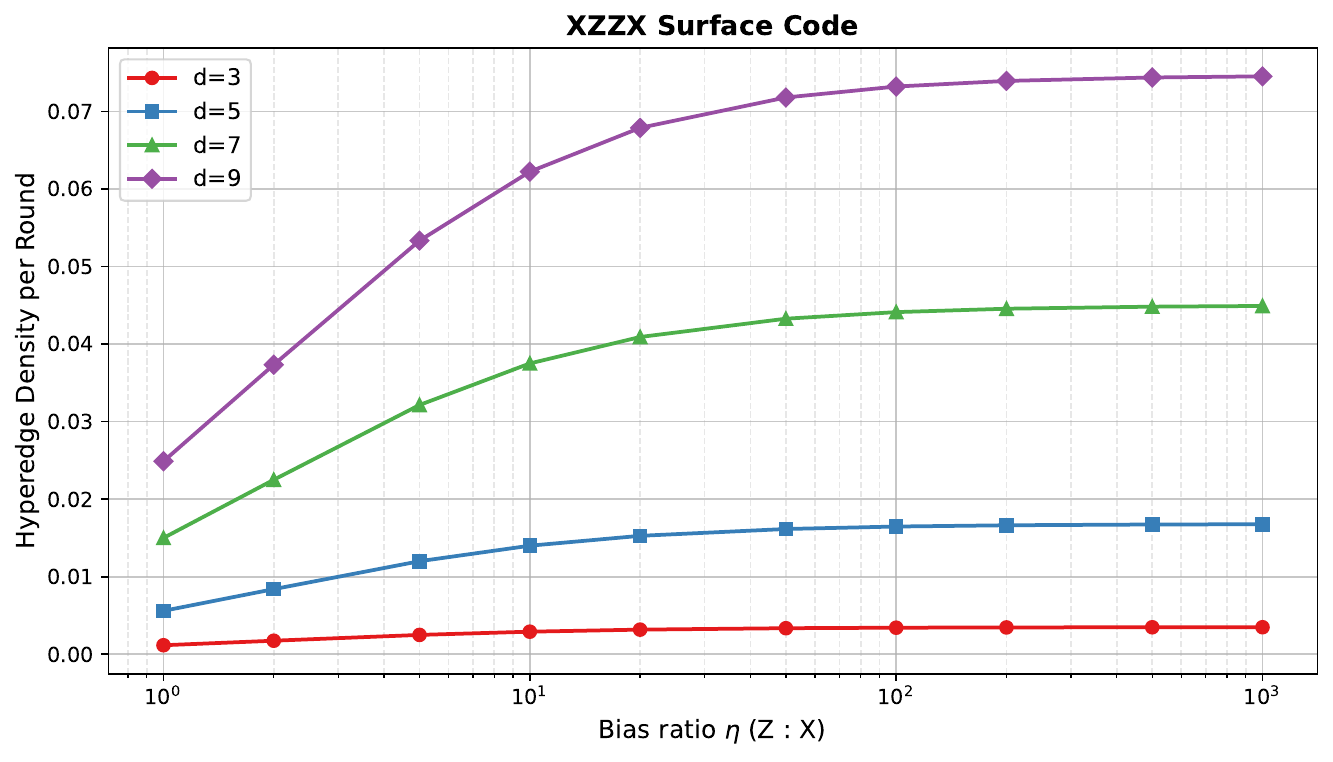}
    \caption{Total hyperedge density per syndrome round as a function of the Z-bias ratio ($\eta$) for the  $\mathrm{XZZX}$ surface code under circuit-level spatio-temporal noise ($d \in \{3, 5, 7, 9\}$, $p=0.007$, rounds $=d$). For every distance, the density of hyperedges monotonically increases and asymptotes at high bias due to the dominance of Z-type spatial correlations, with larger $d$ producing a proportionally higher density owing to the increased number of stabilizers. The threshold minimum observed in Fig. \ref{fig:threshold_vs_eta} at $\eta \approx 10$ occurs despite a lower overall hyperedge density at every distance, highlighting that the code's vulnerability is driven by the specific mixture of Z-hyperedges and transverse X/Y errors rather than absolute hyperedge volume.}
    \label{fig:hyperedge_density}
\end{figure*}

When decoded using the MWPF decoder, the $\mathrm{XZZX}$ surface code continues to outperform the rotated code. For the comprehensive circuit-level spatio-temporal noise model considered, a threshold of approximately $p_{th} \approx 0.70\%$ is observed. 

To properly quantify this performance, these results are compared against their respective pure biased noise baselines. Under ideal phenomenological biased noise, the $\mathrm{XZZX}$ code exhibits a threshold of 7.40\%. The transition to the phenomenological spatio-temporal model introduces a severe degradation of over 74\%, reducing the threshold to $p_{th} \approx 1.86\%$. A similar impact is observed at the circuit-level, where the baseline biased noise threshold of $p_{th} \approx 0.89\%$ falls to the observed 0.70\% when spatio-temporal correlations are introduced. This shift is a direct consequence of the temporal error propagation and faulty gate noise described previously, as continuous dephasing and faulty entangling operations generate correlations that span both space and time. While the circuit-level threshold might initially appear as an improvement over the rotated code, this relative advantage is lost when considering that the underlying noise environment remains heavily dominated by $\mathrm{Z}$ errors.

As a result of this coupling, the one-dimensional symmetry that causes the high threshold of the $\mathrm{XZZX}$ code is fundamentally weakened. Because the $\mathrm{XZZX}$ surface code natively relies on these decoupled 1D structures to protect logical information, it is highly sensitive to multi-basis cross-chain hyperedges. While pure Z-type hyperedges can be efficiently absorbed by the 1D chains, the presence of transverse (X and Y) faults in the spatio-temporal noise actively bridges these chains. This breaks the symmetry that fundamentally underpins the error-correction topological structure, forcing the lattice to behave two-dimensionally and significantly reducing its practical advantage under modeled spatio-temporal noise conditions.

\subsection{Threshold Dependence on Noise Bias}

To fully understand the relationship between spatio-temporal correlations and code geometry, code performance was evaluated under various bias ratios, $\eta \in \{1, 10, 20, 50, 100, \infty\}$. As illustrated in Fig. \ref{fig:threshold_vs_eta}, the two codes respond fundamentally differently to increasing bias under spatio-temporal noise. 

The rotated surface code exhibits a monotonic threshold degradation, decreasing from $p_{th} \approx 0.54\%$ under standard depolarizing noise ($\eta=1$) to $p_{th} \approx 0.34\%$ under pure dephasing ($\eta=\infty$). This predictable decline reflects the structural symmetry of the code; since it does not natively target $\mathrm{Z}$-errors, increasing their proportion uniformly stresses the topological structure alongside the spatio-temporal hyperedges.

In contrast, the $\mathrm{XZZX}$ surface code demonstrates a distinct non-monotonic dependence. Its threshold initially drops from $p_{th} \approx 0.71\%$ at $\eta=1$ to a local minimum of $p_{th} \approx 0.66\%$ near $\eta=10$. However, as the noise becomes overwhelmingly Z-biased, the threshold recovers, climbing asymptotically to $p_{th} \approx 0.70\%$ as $\eta \to \infty$. 

The non-monotonic threshold dependence observed in the $\mathrm{XZZX}$ code can be quantitatively understood by comparing the total hyperedge density against the Pauli error composition. As demonstrated in Fig. \ref{fig:hyperedge_density}, the total density of hyperedges per syndrome round monotonically increases and asymptotes with $\eta$, driven by the dominant Z-type spatial correlations. However, at extreme biases ($\eta \to \infty$), these dense hyperedges are almost exclusively Z-type. Because the MWPF decoder natively processes these hypergraphs without edge decomposition, the $\mathrm{XZZX}$ code successfully absorbs them within its robust, decoupled 1D repetition chains, allowing the threshold to recover.

On the other hand, at moderate biases near $\eta \approx 10$, while the overall hyperedge density is lower, the noise contains a critical fraction of non-dominant transverse errors (X and Y). Since $Y = iXZ$, a transverse Y-fault simultaneously anticommutes with both an X-type and a Z-type stabilizer, flipping detectors of both types in a single event; together with the rarer independent X-faults, these transverse components actively bridge the 1D diagonal strings across different bases, effectively coupling the chains and breaking the one-dimensional decoupling symmetry. This complex, multi-basis hypergraph topology is highly destructive to the $\mathrm{XZZX}$ geometry, resulting in the observed threshold minimum before the purely Z-biased symmetry is restored.

Notably, this behavior is not simply explained by the transverse error fraction alone. The individual transverse error rates $p_x = p_y = p/(\eta+2)$ decrease monotonically with $\eta$, so the transverse fraction is in fact \emph{largest} at $\eta=1$ rather than at the observed minimum near $\eta \approx 10$. The non-monotonic threshold instead arises from a competition between two quantities that individually vary with $\eta$ in opposite directions: the total hyperedge density, which is low at $\eta=1$ and grows with bias (Fig.~\ref{fig:hyperedge_density}), and the transverse error fraction, which is high at $\eta=1$ and falls off with bias. At $\eta=1$, transverse errors are plentiful but too few hyperedges form overall for cross-chain bridging to significantly degrade the threshold. At $\eta \to \infty$, hyperedges are dense but almost purely Z-type, so despite the high density there are too few transverse faults available to bridge diagonal chains, and the code's native resilience to Z-type hyperedges dominates. The intermediate regime near $\eta \approx 10$ is where both conditions are simultaneously satisfied to a damaging degree: hyperedge density has risen enough to generate frequent multi-detector correlations, while the transverse fraction remains large enough to seed cross-chain bridging events. This joint condition, rather than either quantity in isolation, produces the observed threshold minimum.

\section{Discussion and Summary}\label{sec:summary}
The central result of this study is that the effect of increasing Z-bias on the  $\mathrm{XZZX}$ threshold is non-monotonic under spatio-temporal correlated noise. Using the MWPF decoder to natively process correlated syndromes without artificial edge decomposition, we isolated the true topological response of the rotated and  $\mathrm{XZZX}$ surface codes.

While the rotated code exhibits a predictable, monotonic threshold degradation as noise bias increases, the  $\mathrm{XZZX}$ code's one-dimensional decoupling symmetry is actively broken at moderate biases ($\eta \approx 10$). We demonstrated that it is the mixture of transverse errors with spatio-temporal correlations---rather than just the absolute volume of hyperedges---that couples the otherwise independent diagonal repetition strings. Despite this vulnerability, the  $\mathrm{XZZX}$ code retains a higher simulated threshold than the rotated code across the modeled regimes.

It is important to note the limitations of this numerical study. The correlated-noise parameters used ($f_{\text{adj}}$, $f_{\text{diag}}$, $f_{\text{idle}}$) represent specific model assumptions designed to stress the topological structures, rather than direct characterizations of a specific physical hardware platform. Furthermore, while the MWPF decoder effectively bounded the asymptotic code behavior, larger code distances and experimental validation on physical superconducting architectures remain necessary to confirm these thresholds.

Future work should explore bias-preserving gate designs and noise-aware decoding strategies to better maintain the  $\mathrm{XZZX}$ advantage under physical hardware conditions, especially as systems scale below surface code thresholds \cite{google2025below, orourke2025rotated}.

\section*{Code Availability}
The simulation framework and data required to reproduce the results presented in this work are open-source and available on \href{https://github.com/Dhruvkoul1/Comparing-thresholds-of-XZZX-and-rotated-surface-codes-under-spatio-temporal-correlated-noise}{GitHub}.

 \section{Acknowledgment}
 
 The authors are grateful to IISER Bhopal for providing access to the High Performance Computing facility.
 Dhruv Koul is thankful to Mainak Bhattacharyya for his valuable insights and to the members of the QuCIS Lab for their constructive discussions regarding manuscript preparation.

\printbibliography

\end{document}